\documentclass[format=acmsmall, review=false, screen=true]{acmart}

\usepackage{booktabs} 
\usepackage{subcaption}
\usepackage{graphicx}
\usepackage{hyperref}

\usepackage[ruled]{algorithm2e} 

\SetAlFnt{\small}
\SetAlCapFnt{\small}
\SetAlCapNameFnt{\small}
\SetAlCapHSkip{0pt}
\IncMargin{-\parindent}

\acmJournal{TIST}
\acmVolume{9}
\acmNumber{4}
\acmArticle{39}
\acmYear{2017}
\acmMonth{6}
\copyrightyear{2017}

\setcopyright{acmlicensed}

\acmDOI{0000001.0000001}

\begin{document}
\title[The Effect of Pets on Happiness]{The Effect of Pets on Happiness: A Large-scale Multi-Factor Analysis using Social Multimedia}  
\author{Xuefeng Peng}
\affiliation{%
  \institution{University of Rochester}
  \department{Computer Science Department}
  \city{Rochester}
  \state{NY}
  \postcode{14627}
  \country{USA}}
\author{Li-Kai Chi}
\affiliation{%
  \institution{University of Rochester}
  \department{Data Science Department}
  \city{Rochester}
  \state{NY}
  \postcode{14627}
  \country{USA}
}
\author{Jiebo Luo}
\affiliation{%
  \institution{University of Rochester}
  \department{Computer Science Department}
  \city{Rochester}
  \postcode{14627}
  \country{USA}
  }

\begin{abstract}
From reducing stress and loneliness, to boosting productivity and overall well-being, pets are believed to play a significant role in people's daily lives. Many traditional studies have identified that frequent interactions with pets could make individuals become healthier and more optimistic, and ultimately enjoy a happier life. However, most of those studies are not only restricted in scale, but also may carry biases by using subjective self-reports, interviews, and questionnaires as the major approaches. In this paper, we leverage large-scale data collected from social media and the state-of-the-art deep learning technologies to study this phenomenon in depth and breadth. Our study includes four major steps: 1) collecting timeline posts from around 20,000 Instagram users; 2) using face detection and recognition on 2-million photos to infer users' demographics, relationship status, and whether having children, 3) analyzing a user's degree of happiness based on images and captions via smiling classification and textual sentiment analysis; 3) applying transfer learning techniques to retrain the final layer of the Inception v3 model for pet classification; and 4) analyzing the effects of pets on happiness in terms of multiple factors of user demographics. Our main results have demonstrated the efficacy of our proposed method with many new insights. We believe this method is also applicable to other domains as a scalable, efficient, and effective methodology for modeling and analyzing social behaviors and psychological well-being. In addition, to facilitate the research involving human faces, we also release our dataset of 700K analyzed faces. 
\end{abstract}

%

\ccsdesc[500]{Human-centered computing~Social media}

%
%

\keywords{Happiness analysis, happiness, user demographics, pet and happiness, social multimedia, social media.}

\thanks{We thank the support of New York State through the Goergen Institute for Data Science, our corporate research sponsors Xerox and VisualDX, and NSF Award \#1704309.  

  Author's addresses: X. Peng, Computer Science Department, University of Rochester; L. Chi, Data Science Department, University of Rochester {and} J. Luo, Computer Science
  Department, University of Rochester.}

\maketitle

\renewcommand{\shortauthors}{X. Peng et al.}

\section{Introduction}

Happiness has always been a subjective and multidimensional matter; its definition varies individually, and the factors impacting our feeling of happiness are diverse. A study in \cite{ura_2016} has constructed a mechanism for indexing people's happiness through considering nine domains, namely psychological wellbeing, time use, community vitality, cultural diversity, ecological resilience, living standard, health, education, and good governance. Our study focuses on analyzing happiness from the perspectives of psychological wellbeing. It has shown by \cite{Glenn1988} and \cite{Budge1998}, satisfactory companion relationship is a crucial factor towards our wellbeing; the companionship could be between human and human as well as human and animal; and those relationships often interweave with each other. In this paper, we are particularly interested in analyzing how pet companionship, relationship status (i.e. having partner or not), and having children affect happiness though a data-driven approach.

\begin{figure}
  \centering	
  \includegraphics[width=\textwidth]{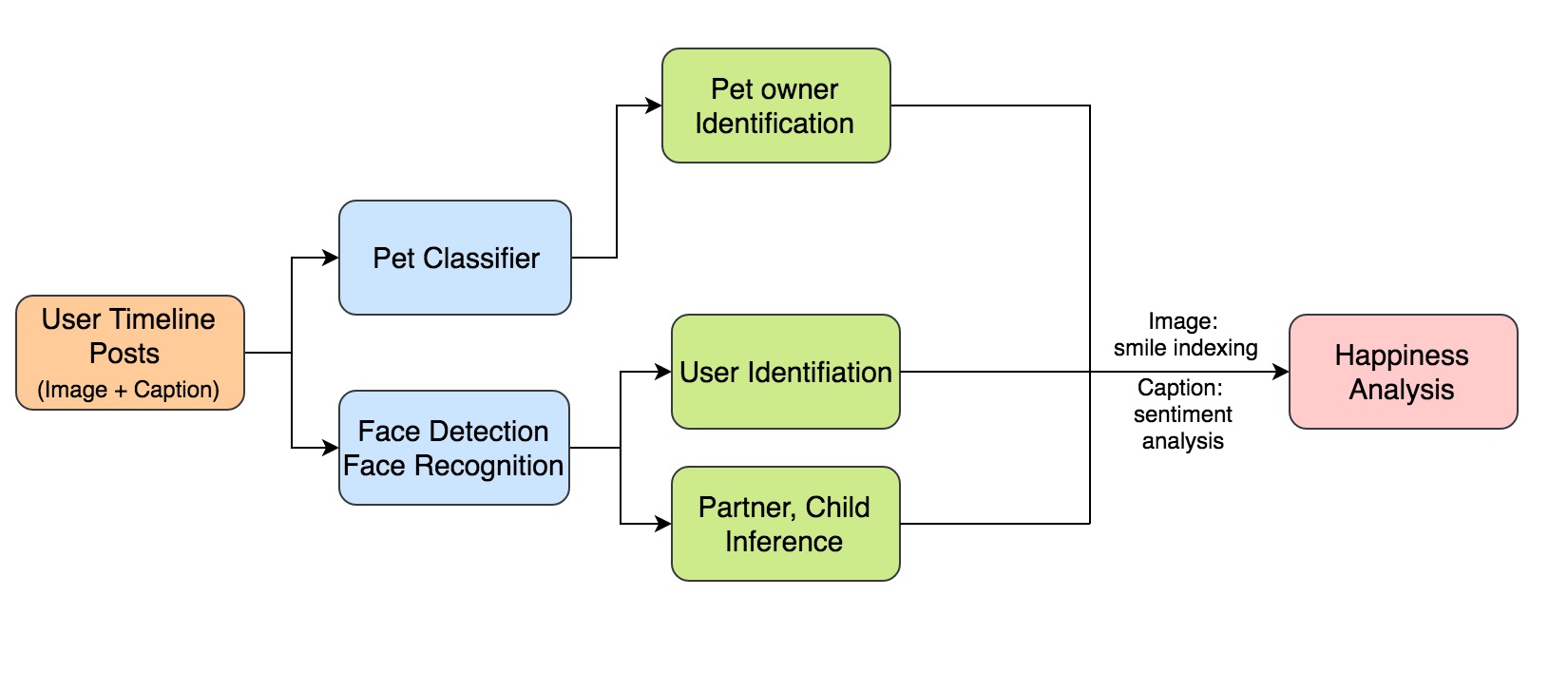}
  \caption{The proposed computational framework for multi-factor user happiness analysis. Given user timeline posts, the pet classifier first identifies if he/she is a pet owner; then the state-of-the-art facial detection and recognition technologies provided by Face++ are applied on all the posts with faces for user demographics inference; user relationship status, and if having children inferences; and finally, user happiness analysis.}
  \label{fig:one}
\end{figure}

Our study proposes a novel, efficient, and scalable computational framework, shown in Figure \ref{fig:one}, to assess the effects of pets, relationship status, and having children on individual happiness. More specifically, for each user obtained from Instagram, we first apply our pet classifier to identify if he/she is a pet owner; then we use the state-of-the-art facial detection and recognition technologies provided by Face++ on all the posts with faces for 1) inferring user demographics; 2) inferring user relationship status, and if the user has children; and 3) analyzing user happiness.

Our contributions can be summarized in fourfold as follows:

\begin{itemize}
\item We apply the cutting-edge transfer learning to construct an extremely high-performance pet classifier retrained specifically on social media posts.
\item We propose an efficient way to infer user's relationship status, and if having children.
\item We analyze the effects of pets on happiness, and further analyze the relationship status, and having children on happiness in pet and none-pet owner groups, respectively. 
\item We further show the effectiveness and efficiencies of combining social media resources and computational methods on large-scale screening and analysis of user-level behaviors, and psychological well-being.
\item To help other research involving human faces, we share our dataset\footnote{\label{repolink}\url{https://github.com/xuefeng7/FACE-LIBRARY} }including nearly 700K processed human faces from social media posts.
\end{itemize}

\section{RELATED WORK}

Several previous studies have demonstrated the positive impacts possessing pets for elderly \cite{Garrity1989,Ory1983}, while our study focuses on further confirming this effect using a large number of samples from social media regardless of age, gender, and ethnicity. 

Our study is inspired by the preliminary research in \cite{Wu2016} but we extend the analyses in four distinguishable ways: 1) using nearly ten-times more data, 2) retraining the final layer of the Inception model v3 as our pet classifier, which achieves a superior accuracy, 3) improving the method of identifying user faces in user timeline posts, and most importantly, 4) considering {\it multiple factors} including user's relationship status and if having children. In contrast, the work in \cite{Wu2016}  analyzed 300,000 posts from around 2900 users, while our study analyzes around 2-million posts from roughly 20,000 Instagram users. \cite{Wu2016} has constructed a reasonable convolutional neural network (CNN) as its pet classifier while we build a high-performance classifier by retraining the final layer of the Inception v3 model using remarkably fewer training samples. In addition, \cite{Wu2016} assumed that the largest face in a selfie post is the face of the user, and analyzed the average happiness of this user according to all of those "largest" faces throughout this user's timeline posts. While this is usually valid as most likely the user would be the one who is holding the camera when taking selfies,  it is still possible that the largest face is not from the user. We reduce such inaccuracy by employing face recognition to identify user faces. Lastly but most importantly, our work extensively analyzes the multi-factor effects of relationship status, and if having children on happiness among pet and none-pet owners.

\section{DATA ACQUISITION}

With billions of users, Instagram is a rich source of high quality, and keyword tagged images. In our study, timeline posts from both potential pet and none-pet owners are needed. For potential pet owners, considering that dogs and cats are the two most common pets all over the world, we collected users who have posted either dog or cat images by retrieving posts tagged with dog or cat related keywords. To reduce sample biases, multiple hashtags were considered, they are \#mydog, \#mypuppy, \#mydoggie, and \#mycat, \#mykitten, \#mykitty for cats and dogs, respectively. After obtaining the usernames, we backtracked up to 100 timeline posts from each user for later analysis. None-pet owners were obtained in a similar fashion, except the hashtags included are \#selfies, \#me, and \#life. Eventually, we collected nearly 20,000 users and 2-million posts from their timelines. 

Human face analysis is becoming increasingly popular in the fields of computer vision and artificial intelligence. In order to facilitate related research, we have decided to publish our dataset whose web link can be found in  Footnote \ref{repolink} . The dataset contains 700K processed faces that originated from social media posts. Facial attributes such as age, gender, race, as well as facial landmarks for each face, are provided.

\section{PET OWNER IDENTIFICATION}

Two steps are taken to identify if an Instagram user is a pet owner or not. First, we classify each image in a user's timeline into three classes, namely dog, cat, and others. Next, since a none-pet owner might post pet images while visiting a friend who is a pet owner, we identify a pet owner by looking at the frequency of posting pet image throughout the timeline. We describe the details in the following subsections.

\subsection{Classification}

\subsubsection{Inception v3 Model}
Building a robust deep learning model for object recognition can cost enormous amount of computing power and time; therefore, we chose to develop our pet classification model using the state-of-the-art transfer learning \cite{Weiss2016}. This technique allows us to take advantage of a fully-trained model, and retrain its final layer for new categories with far less training samples. The fully-trained model we adopt is the Inception v3 model built by Google. The detailed structure of the model can be found in \cite{Szegedy2016}, and according to \cite{Szegedy2016}, Inception v3 outperforms many other state-of-the-art deep convolutional networks such as VGGNet, GoogleNet, BN-Inception, and so on, on the ILSVR2014 classification challenge validation set \cite{Russakovsky2015}. 

\subsubsection{Retrain Inception v3 Model}
\begin{figure}
  \begin{subfigure}{0.6\linewidth}
  \includegraphics[width=.25\linewidth]{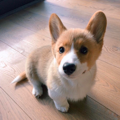}\hfill
  \includegraphics[width=.25\linewidth]{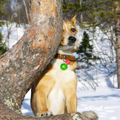}\hfill
  \includegraphics[width=.25\linewidth]{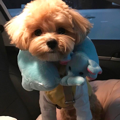}\hfill
  \includegraphics[width=.25\linewidth]{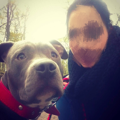}
  \caption{Dog}
  \end{subfigure}\par\medskip
  \begin{subfigure}{0.6\linewidth}
  \includegraphics[width=.25\linewidth]{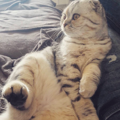}\hfill
  \includegraphics[width=.25\linewidth]{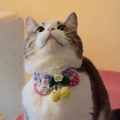}\hfill
  \includegraphics[width=.25\linewidth]{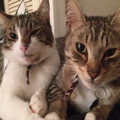}\hfill
  \includegraphics[width=.25\linewidth]{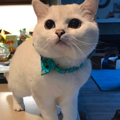}
  \caption{Cat}
  \end{subfigure}\par\medskip
  \begin{subfigure}{0.6\linewidth}
  \includegraphics[width=.25\linewidth]{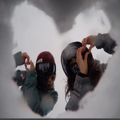}\hfill
  \includegraphics[width=.25\linewidth]{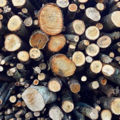}\hfill
  \includegraphics[width=.25\linewidth]{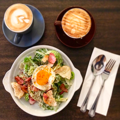}\hfill
  \includegraphics[width=.25\linewidth]{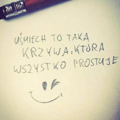}
  \caption{Others}
  \end{subfigure}
  \caption{Sample training images of three categories. We train the model by samples specifically collected from Instagram so that the model can produce better prediction performance on the large-scale Instagram dataset we used for analysis.}
  \label{fig:two}
\end{figure}

We retrain the final output layer of the Inception v3 model for three new categories: a) dog, b) cat, and c) others. We collected only 2000 images of each category from Instagram, and manually labeled them as our training set. The dog and cat images contain at least one dog or cat, and others can contain any scene except animals. Figure \ref{fig:two} shows a few sample images from each category. We utilize Tensorflow as our computation backend; and with a single Nvidia Tesla K20X GPU, the entire retraining process takes approximately 10 hours.

The model is retrained through 600 steps with the 0.1 default initial learning rate. To avoid over-fitting, we split 80\% of the entire dataset as our main training set, 10\% as the testing set, and the remaining 10\% as the validation set. Also, during the retraining phase, 10\% of these training images were randomly cropped, flipped, scaled, mirrored, and brightness-adjusted for improving model adaptability. 

\subsubsection{Retrained Model Validation}
\begin{figure}
  \centering	
  \includegraphics[width=0.8\textwidth]{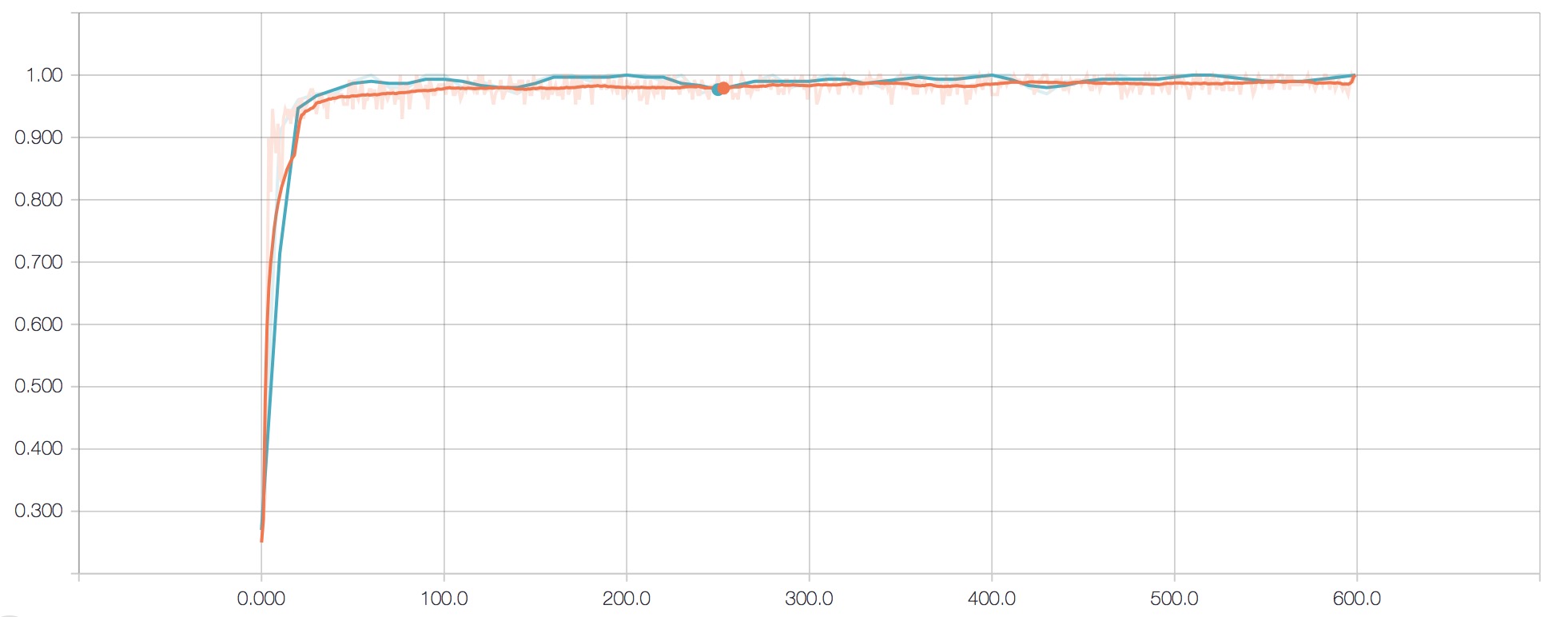}
  \caption{The testing and validating accuracy curves. The testing accuracy curve is marked in orange, and the validating curve is marked in blue.}
  \label{fig:three}
\end{figure}

\begin{figure}
  \centering	
  \includegraphics[width=0.8\textwidth]{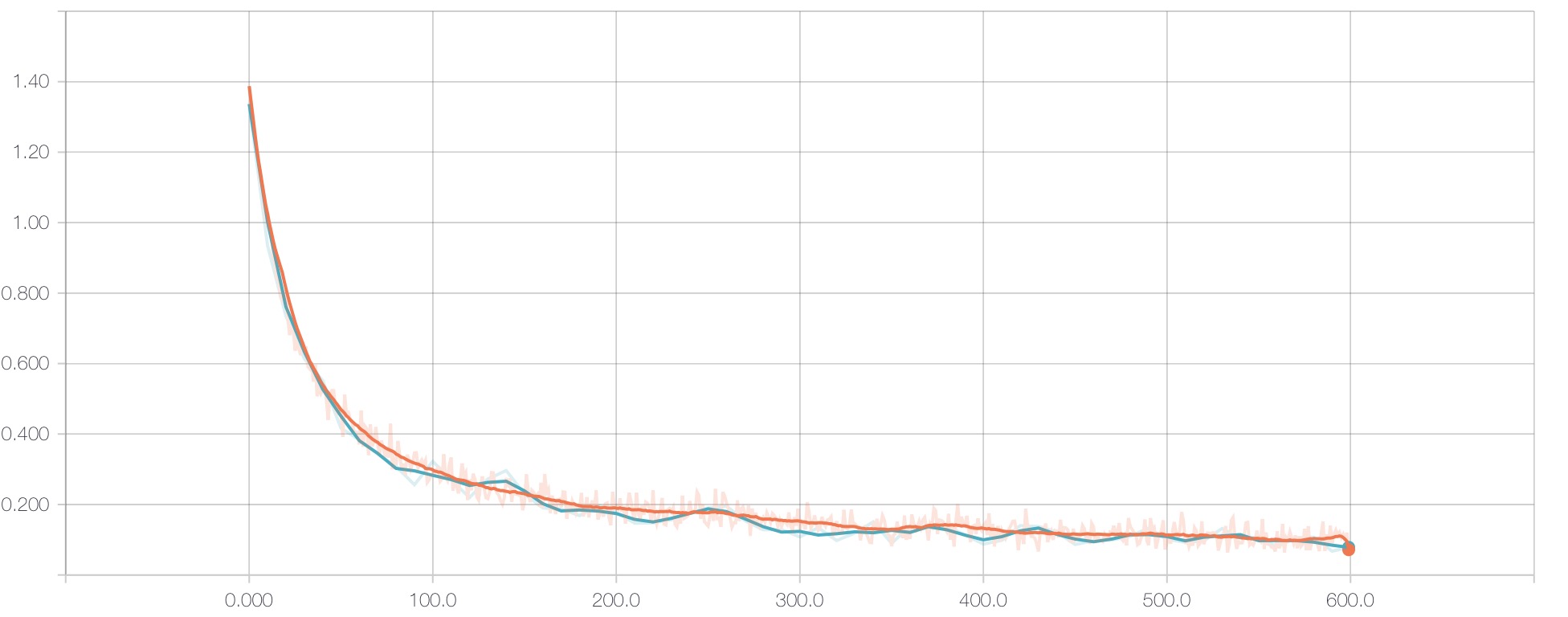}
  \caption{The testing and validating cross-entropy curves. }
  \label{fig:four}
\end{figure}

In the final step, the retrained model achieves 98.4\% testing accuracy, and 100\% validation accuracy. Cross-entropy for testing and validation sets decreased from 0.9268 and 1.051 to 0.04 and 0.03 at the end of retraining, respectively. Figure \ref{fig:three} shows the accuracy curves, and Figure \ref{fig:four} shows the cross-entropy curves.

\begin{figure}
  \centering	
  \includegraphics[width=0.5\textwidth]{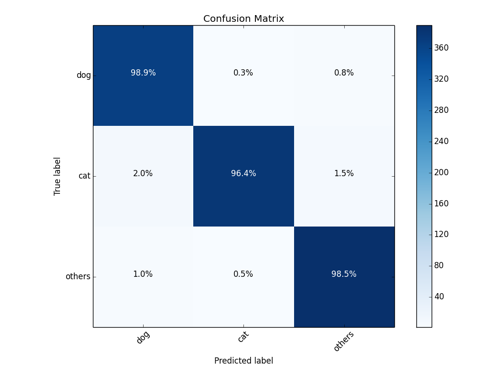}
  \caption{The confusion matrix of the retrained Inception v3 pet classifier}
  \label{fig:five}
\end{figure}

We further verify our model by classifying 1500 manually-labeled unseen images for each class collected from Instagram, the retrained model achieves high accuracy of 99.0\%, 96.4\%, and 98.5\% for dog, cat, and others, respectively. The confusion matrix is shown in Figure \ref{fig:five}. 

\subsection{Identification}

We use two criteria to identify if a user is a pet owner. First, we collect users who have post images tagged with either dog or cat related topic as described in section 3. Tags such as \#mydog, \#mycat, etc. are strong indicators that these users own pets; however, exceptions may exist. Therefore, we develop the pet classifier to further assure that there are indeed dogs or cats in a pet owner's timeline. Furthermore and more importantly, since a none-pet owner may post pet images while visiting a friend who is a pet owner, we also took the frequency of posting pet images into consideration. We treat one week as a time window, and a user is considered as a pet owner only if the user posts images of the same type of pet, namely dog or cat, more than one time window throughout his/her timeline. If a user post pet images only within one time window (one week), this person is not considered as a pet owner regardless the amount of pet pictures post. This is similar to the approach adopted in \cite{Wu2016}. 

\section{TIMELINE ANALYSIS}
Face++ is an open source face engine with both an online  API and an offline SDK that provide services including face detection, face recognition and face analysis. The  system is built with a CNN structure similar to the ImageNet structure as discussed in section IV-A \cite{krizhevsky2012imagenet}, where five convolutional layers with maxpooling are connected with two fully connected layers and a softmax layer on top of them. 

In addition, Face++'s services have been widely used in commercial applications by many major companies in China. By integrating Face++, AliPay, used by more than 120 million people, allows users to transfer money securely by using their faces as credentials. Meanwhile, Didi, China's dominant ride-hailing company, uses the Face++ software to allow passengers to confirm the driver's identity. Lenovo, a world-wide computer manufacturer, also adopts the biologic identification solution from Face++.

We adopt the reliable services provided by the Face++ engine \cite{Fan2014} to detect, analyze, and recognize faces in each user's timeline for three purposes: 
\begin{itemize}
\item Identify a user's faces and infer his/her demographics such as age, gender, and race.
\item Identify the frequently appeared faces in user's timeline, and infer their demographics.
\item Analyze the smiles of user faces to infer his/her overall happiness.
\end{itemize}

\subsection{Face detection}

Given a user, face detection is applied to every image within his/her timeline. Note that if a user's timeline contains less than five faces, such user will be discarded from further analysis. Besides, since more posts will assist us to identify more accurately if user is a pet owner, or if a user has a partner, we also discard user whose timeline contains less than 25 posts from further analysis.

\begin{figure}
  \begin{subfigure}{0.2\linewidth}
  \centering
  \includegraphics[width=.9\linewidth]{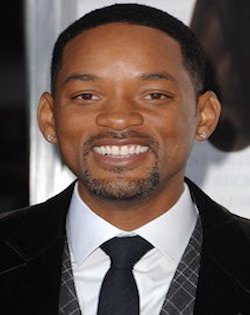}\hfill
  \caption{Age: 39\\\hspace{\textwidth}Gender: Male\\\hspace{\textwidth}Smiling: 54.10\\\hspace{\textwidth}Race: Black}
  \end{subfigure}
  \begin{subfigure}{0.2\linewidth}
  \centering
  \includegraphics[width=.9\linewidth]{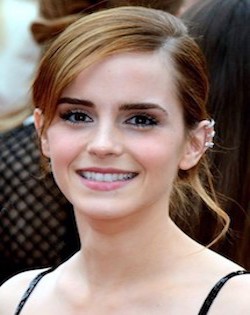}\hfill
  \caption{Age: 12\\\hspace{\textwidth}Gender: Female\\\hspace{\textwidth}Smiling: 94.68\\\hspace{\textwidth}Race: white}
  \end{subfigure}
  \begin{subfigure}{0.2\linewidth}
  \centering
  \includegraphics[width=.9\linewidth]{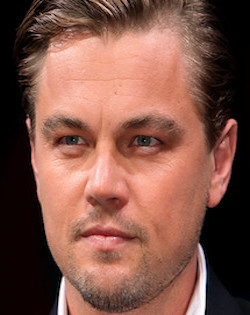}\hfill
  \caption{Age: 41\\\hspace{\textwidth}Gender: Male\\\hspace{\textwidth}Smiling: 1.20\\\hspace{\textwidth}Race: white}
  \end{subfigure}
   \begin{subfigure}{0.2\linewidth}
  \centering
  \includegraphics[width=.9\linewidth]{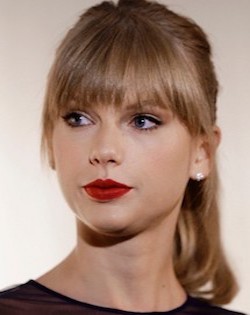}\hfill
  \caption{Age: 33\\\hspace{\textwidth}Gender: Female\\\hspace{\textwidth}Smiling: 2.63\\\hspace{\textwidth}Race: white}
  \end{subfigure}
  \caption{Examples of demographics and smiling attribute extracted from detected faces.}
  \label{fig:six}
\end{figure}

For each detected face, its attributes including gender, age, race, smiling, glass and pose will be computed. The attributes we are interested in are age, gender, race, and smiling. Figure \ref{fig:six} shows few examples of applying Face++ on faces. 

Face++ uses the same methodology developed in \cite{Kumar2009} to detect smiling. Note that, as expected, the smiling values of the faces in Figures \ref{fig:six} (a) and (b) are much higher than these in Figures \ref{fig:six} (d) and (c). This score is an important indicator for our later happiness analysis. 

\subsection{Face grouping}
It is possible that photos in a user's timeline contain not only his/her own faces, but also the faces of his/her friends, family members, and even strangers. Therefore, we need to group the faces into several face sets where each face set consists of only one individual's faces. To achieve this, for each user, 1) we create a faceset, 2) detect every face in user's timeline posts, and add them into the faceset, 3) utilize face recognition technology \cite{Zhou2013} to group faces in the faceset.

\begin{figure}
  \centering	
  \includegraphics[width=0.8\textwidth]{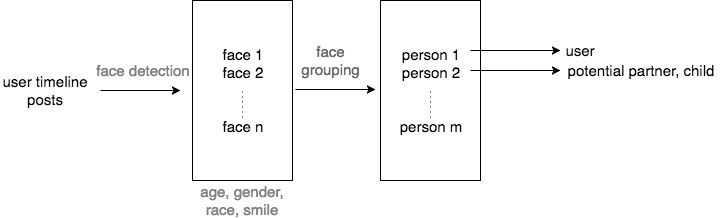}
  \caption{The process of face detection and face grouping for user identification, and inferring relationship status, and if having children. }
  \label{fig:seven}
\end{figure}

Figure \ref{fig:seven} shows the entire process from face detection to face grouping.

\subsubsection{User Identification}
The person list in Figure \ref{fig:seven} is sorted in a descending order by the number of faces. We consider the most frequently appeared face as the user's face. Based on this assumption, in the grouped faceset, we can easily locate the user. We conduct the happiness analysis solely for the faces identified as the user. 

\subsubsection{Relationship Status, Having-Children Inference}
As shown in Figure \ref{fig:seven}, the second and third most frequently appeared faces in a user's timeline may have close relationships with the user. We further infer such relationship by checking the age. Similar to how we identify pet owners, we also consider the frequency while determining if a user has a partner or child. A user is considered to have a partner, if another person's face appears throughout the user's timeline for more than a time window (one week), and the age difference between the user and the person is less than 5 years. In the same vein, a user is considered to have a child if 1) the user is older than 18 years old, and 2) another person's face appears throughout the user's timeline for more than one time window, and the age difference between the user and the person is more than 18 years.

\subsection{Happiness Analysis}
We consider both visual and textual context in our study of happiness. 

\subsubsection{Visual Context}
Our methodology of quantifying user happiness is in the same vein as [4, 17]. Let $H_{t}$ denotes the visual happiness score of a user over a period of $t$, and $|F_{t}|$ denotes the number of user faces in the timeline over the time frame $t$, and $S_{i}$ indicates the confidence score of smiling for user face appeared in the post $i$. Equation \ref{eqn:01} calculates the final visual happiness score for any given user.

\begin{equation}
\label{eqn:01}
H_{t} = \frac{\sum_{i\in t}S_{i}}{|F_{t}|}
\end{equation}

Note that the smiling confidence score for each face is acquired during the face detection phase. 

\subsubsection{Textual Context}
We also utilize the sentiment analysis on the user post captions to infer user happiness. In particular, we take advantage of Valence Aware Dictionary and sEntiment Reasoner (VADER) \cite{Hutto2014} as our lexicon and rule-based sentiment analysis tool. This analyzer is specifically attuned to sentiments expressed over social multimedia. Moreover, VADER is particularly suitable for our study as it can capture many sentiment-laden slangs, emoticons, initialisms, and acronyms such as ":)", ":D", "friggin", "sux", and "lol", which are prevalent in Instagram post captions. According to \cite{Hutto2014}, for each given caption, a normalized, weighted composite score from -1 (most extreme negative) to 1 (most extreme positive) is generated to measure its emotion. In our study, we use the average composite score to infer user happiness. Let $C_{t}$ denotes the textual happiness score of a user over a period of $t$, and $|P_{t}|$ denotes the number of all user captions in the timeline over the time frame $t$, and $W_{i}$ indicates the composite score of the caption for user timeline post $i$. Like Equation 1, Equation \ref{eqn:02} calculates the final textual happiness score.

\begin{equation}
\label{eqn:02}
C_{t} = \frac{\sum_{i\in t}W_{i}}{|P_{t}|}
\end{equation}

\section{MAIN RESULTS}

In this section, we present our findings. We first show the demographic distribution of our users in section 6.1. Then in section 6.2, we investigate the happiness over pet and none-pet owners separately in terms of relationship status, and if having children.

\subsection{Demographics}
In our study, we process over 20,000 users, but the filter step described in section 5.1, and the pet-owner identification step together help us refine our user collection. 10,536 users were kept for our happiness analysis. In terms of gender, we have 7308 females and 3228 males; for race, there are 2210 Asians, 681 African Americans, and 7654 Caucasians. Table \ref{table:one} summarized the gender and race distributions. 

\begin{table}[H]
\centering
\caption{Gender and race distribution,  as shown below, the majority of collected faces are Caucasian, followed by Asian and African American. Moreover, the Female faces are about twice more than male faces.}
\label{table:one}
\begin{tabular}{cccc|c}
\hline
                            & Asian & African American & Caucasian & Sum   \\ \hline
\multicolumn{1}{c|}{Male}   & 562   & 333              & 2333      & 3228  \\
\multicolumn{1}{c|}{Female} & 1648  & 348              & 5312      & 7308  \\ \hline
\multicolumn{1}{c|}{Sum}    & 2210  & 681              & 7645      & 10536 \\ \hline
\end{tabular}
\end{table}

\begin{figure}
  \centering	
  \includegraphics[width=0.8\textwidth]{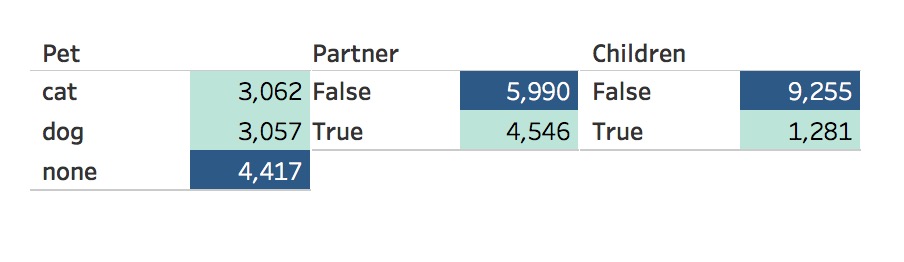}
  \caption{Pet owner, partner, and children distributions. As shown, In terms of Pet and Partner categories, the number of users are balanced; however, for the last category, amount of users without child is much higher than that of users with child. }
  \label{fig:eight}
\end{figure}

Figure \ref{fig:eight} shows the distribution of pet, partner, and children.

\subsection{Happiness}
In this subsection, we present happiness score comparisons. Note that we report $\bar{H_{t}}$ and $\bar{C_{t}}$ in parallel along with the significance tests. The statistical significance test we employed is multiple comparison test \cite{Hochberg1987}, which compares the means of several groups to test the hypothesis that they are all equal, against the general alternative that they are not all equal. This testing procedure also provides the information about which pairs of means are significantly different. Also note we used the 95\% confidence level throughout our analysis.

\subsubsection{Pet and None-pet owners}

\begin{figure}
  \centering	
  \includegraphics[width=0.6\textwidth]{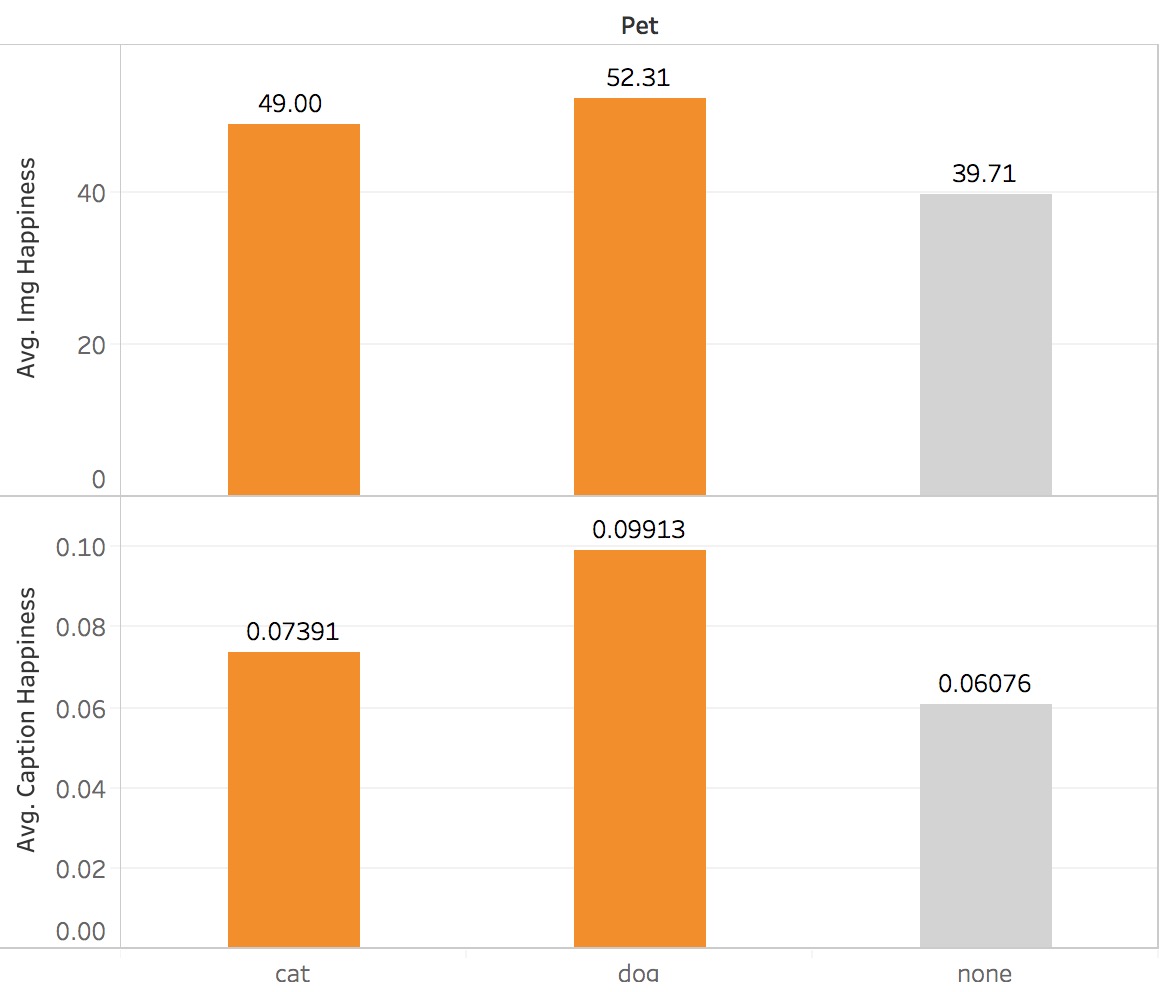}
  \caption{The mean happiness score comparison between dog, cat, and none-pet owners. The upper and lower chart represent $\bar{H_{t}}$ and $\bar{C_{t}}$, respectively. As shown, users with pet, either cat or dog, have higher mean happiness score than non-pet owners when their photos and captions are analyzed.} 

  \label{fig:nine}
\end{figure}

The multiple comparison test results are presented in Table \ref{table:two}.

\begin{table}
\centering
\caption{Multiple comparison test results for happiness scores over pet and none-pet owners.}
\label{table:two}
\begin{tabular}{ccccc}
\hline
\multicolumn{5}{c}{$\bar{H_{t}}$}                                  \\ \hline
categories & lower   & est. mean diff & upper   & p-val \\ \hline
dog-cat    & 1.6018  & 3.3093         & 5.0168  & 0     \\
dog-none   & 10.9853 & 12.5968        & 14.2084 & 0     \\
cat-none   & 7.6883  & 9.2875         & 10.8868 & 0     \\ \hline
\multicolumn{5}{c}{$\bar{C_{t}}$}                                  \\ \hline
dog-cat    & 0.0187  & 0.0252         & 0.0317  & 0     \\
dog-none   & 0.0323  & 0.0384         & 0.0445  & 0     \\
cat-none   & 0.0071  & 0.0132         & 0.0192  & 0     \\ \hline
\end{tabular}
\end{table}

$\bar{H_{t}}$ and $\bar{C_{t}}$ share the same trends among dog, cat, and none-pet owners. Dog owners possess the highest scores, followed by cat owners, and then none-pet owners. If we combine dog and cat owners as pet owners, then the significant test denotes that $\bar{H_{t}}$ and $\bar{C_{t}}$ of pet owner are higher than that of none-pet owner by $10.92 \pm 1.14$, with p-values both equal to zero. 

Note that we merge dog and cat owners as pet owners for the rest of analyses.

\subsubsection{Gender}
We compare the $\bar{H_{t}}$ and $\bar{C_{t}}$ over different genders among pet and none-pet owners.

\begin{figure}
  \centering	
  \includegraphics[width=0.6\textwidth]{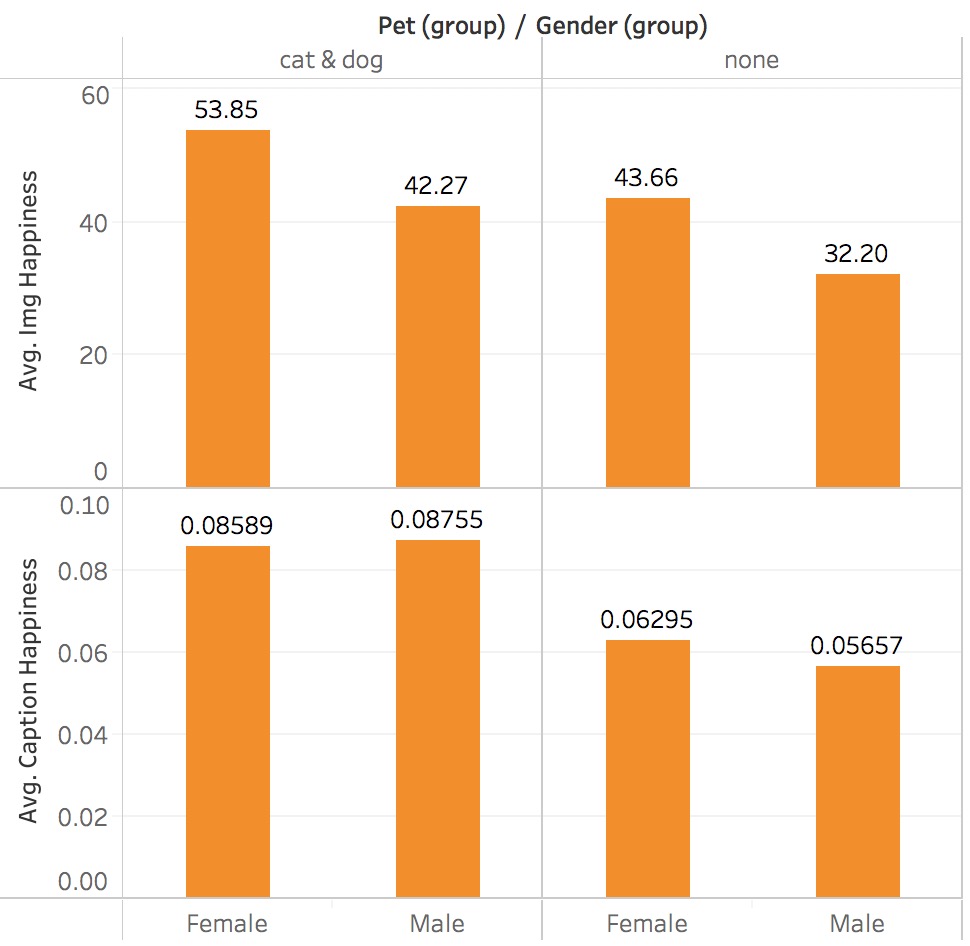}
  \caption{The mean happiness score comparisons between genders among pet and none-pet owners. As shown, both genders show higher mean happiness score when they have pet, and female typically have a higher happiness score than male.
}
  \label{fig:ten}
\end{figure}

The test results suggest that for pet and none-pet owners, the differences of $\bar{H_{t}}$ between female and male are $11.58 \pm 2.05$ (p-value=0), $11.46 \pm 2.4$ (p-value=0), respectively. No significant $\bar{C_{t}}$ differences were found between genders for both groups. 

\subsubsection{Race}
We also compare the $\bar{H_{t}}$ and $\bar{C_{t}}$ over different race groups among pet and none-pet owners.

\begin{figure}
  \centering	
  \includegraphics[width=0.6\textwidth]{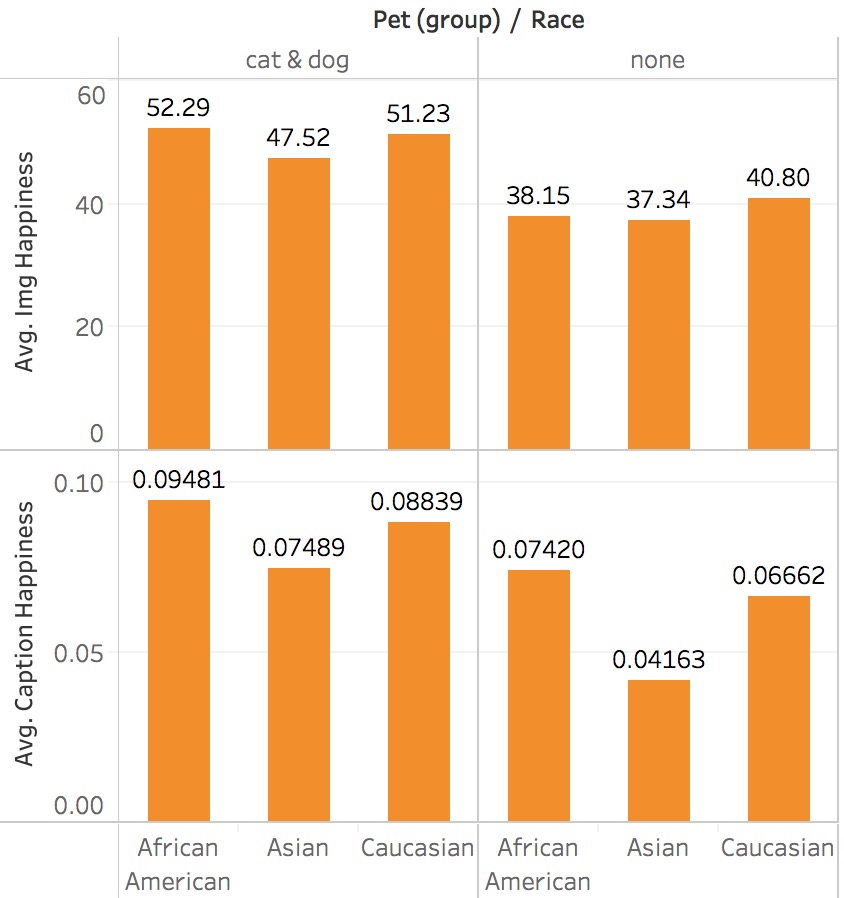}
  \caption{The mean happiness score comparisons between races among pet and none-pet owners. Pet owners have higher average happiness score than none-pet owners regardless of race. It also shows that Asians have a lower average happiness score when compared to African American and Caucasian.}
  \label{fig:eleven}
\end{figure}

In terms of $\bar{H_{t}}$, the difference between Asian and Caucasian pet owners is significant, and its value is $-3.17 \pm 2.74$ (p-value=0.0016). Similarly, among none-pet owners, the $\bar{H_{t}}$ of Asian is $-3.46 \pm 2.98$ (p-value=0.0121) lower than that of Caucasian. The significant differences of $\bar{C_{t}}$ among different racial groups are summarized in Table \ref{table:three}.

\begin{table}
\centering
\caption{Multiple comparison test results for happiness scores over races among pet and none-pet owners.}
\label{table:three}
\begin{tabular}{ccccc}
\hline
\multicolumn{5}{c}{Pet Owner}                                    \\ \hline
categories         & lower   & est. mean diff & upper   & p-val  \\ \hline
Asian - Caucasian  & -0.0239 & -0.0135        & -0.0031 & 0.0029 \\
Asian - African. A & -0.0392 & -0.0199        & -0.0006 & 0.0391 \\ \hline
\multicolumn{5}{c}{None Pet Owner}                               \\ \hline
Asian - Caucasian  & -0.0363 & -0.0250        & -0.0137 & 0      \\
Asian - African. A & -0.0527 & -0.0326        & -0.0124 & 0.001  \\ \hline
\end{tabular}
\end{table}

In general, in both pet and none-pet groups, Asian has lower $\bar{H_{t}}$ and $\bar{C_{t}}$ than those of Caucasian and African Americans.

\subsubsection{Partner}
In this subsection, we investigate the effect of having a partner on happiness for pet and none-pet owners.

\begin{figure}
  \centering	
  \includegraphics[width=0.6\textwidth]{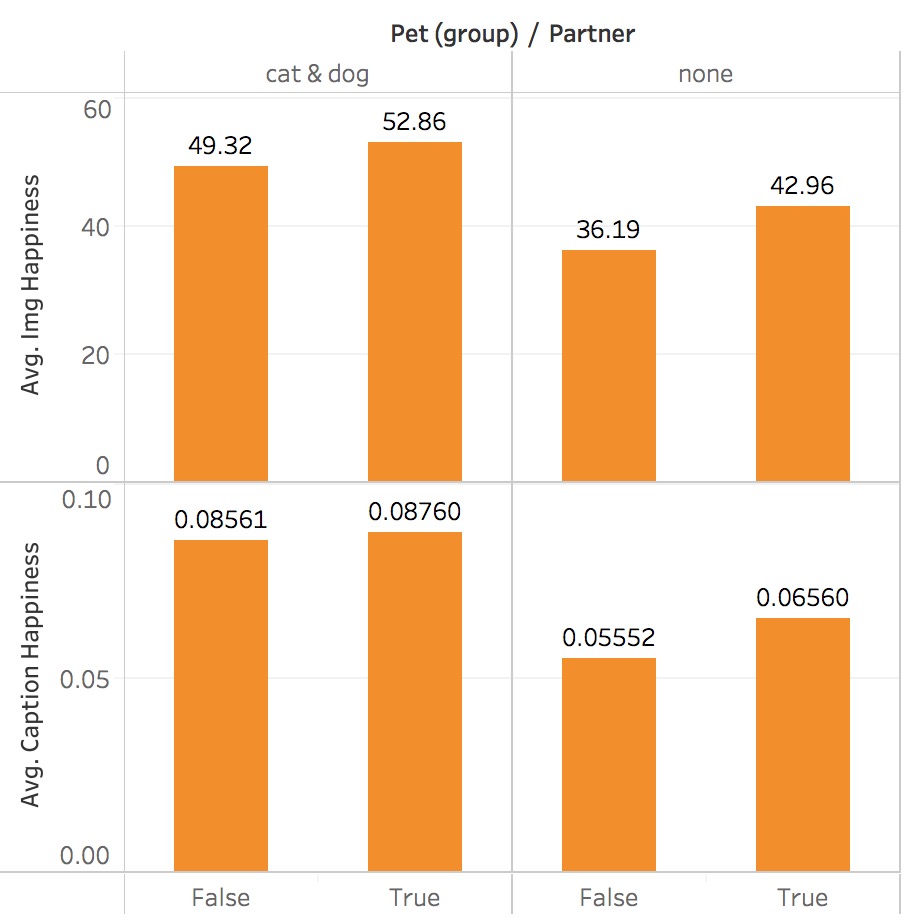}
  \caption{The mean happiness score comparisons between having partner and having no partner among pet and none-pet owners. According to the chart, pet owners have higher average happiness score than none-pet owners, and people with partners have higher average happiness score than people without partners.}
  \label{fig:twelve}
\end{figure}

As shown in the Figure \ref{fig:twelve}, $\bar{H_{t}}$ and $\bar{C_{t}}$ share the same trends over population having partner and having no partner in both pet and none-pet owner groups. More specifically, in both groups, those who have partners possess higher $\bar{H_{t}}$ and $\bar{C_{t}}$. The significant tests suggest that among pet owners, $\bar{H_{t}}$ for those who have partner are $3.54 \pm 1.93$ (p-value=0.0) higher than that for those who have no partner. No statistically significant difference exists with respect to $\bar{C_{t}}$. 

Among none-pet owners, $\bar{H_{t}}$ and $\bar{C_{t}}$ for those who have partner are $6.77 \pm 2.31$ (p-value=0.0), and $0.01 \pm 0.0089$ (p-value=0.0) higher. 

\subsubsection{Child}
In this subsection, we examine the effect of having children on happiness for pet and none-pet owners.

\begin{figure}
  \centering	
  \includegraphics[width=0.6\textwidth]{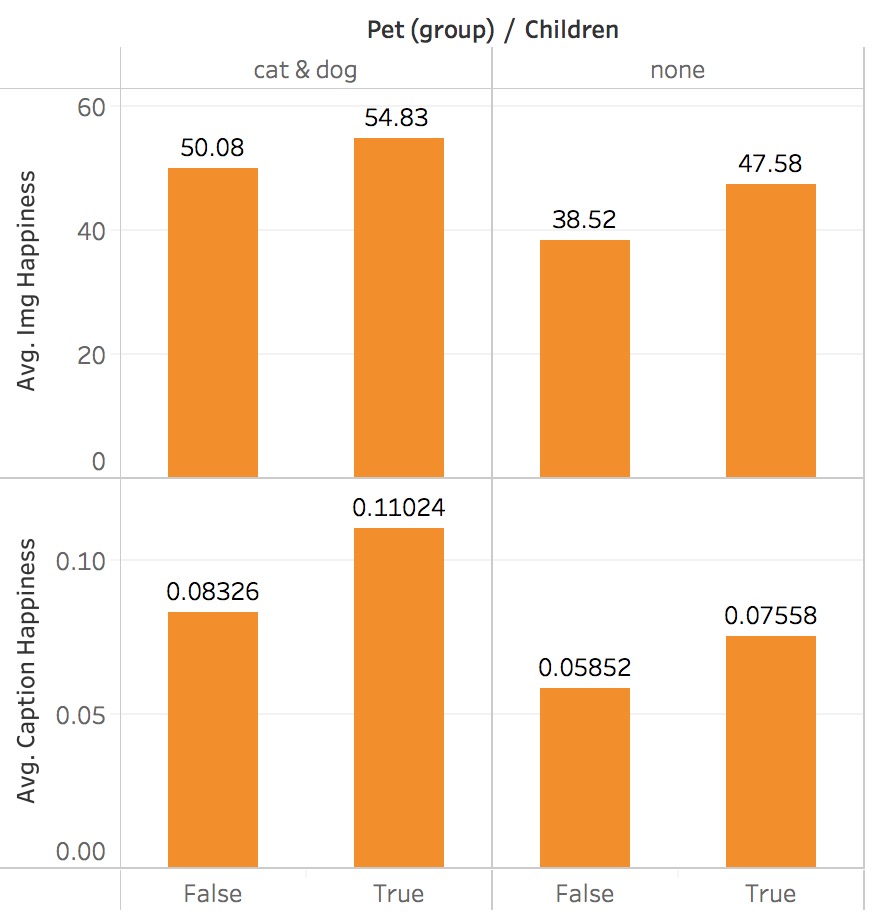}
  \caption{The mean happiness score comparisons between having child and having no child among pet and none-pet owners. The chart illustrates that pet owners have higher average happiness score than none-pet owners, and people with children have higher average happiness score than people without children.}
  \label{fig:thirdteen}
\end{figure}

As Figure \ref{fig:thirdteen} shows, in both pet and none-pet owner groups, the population with children possess higher $\bar{H_{t}}$ and $\bar{C_{t}}$. We present the significant test results in Table \ref{table:four}.

\begin{table}
\centering
\caption{Multiple comparison test results for happiness scores over having child and having no child population among pet and none-pet owners.}
\label{table:four}
\begin{tabular}{ccccc}
\hline
\multicolumn{5}{c}{Pet Owner}                      \\ \hline
score & lower  & est. mean diff & upper   & p-val  \\ \hline
$\bar{H_{t}}$    & 1.8128 & 4.7442         & 7.6756  & 0.0002 \\
$\bar{C_{t}}$    & 0.0158 & 0.0270         & 0.0381  & 0      \\ \hline
\multicolumn{5}{c}{None Pet Owner}                 \\ \hline
$\bar{H_{t}}$    & 5.6261 & 9.0561         & 12.4861 & 0      \\
$\bar{C_{t}}$    & 0.004  & 0.0171         & 0.0301  & 0      \\ \hline
\end{tabular}
\end{table}

\section{CONCLUSIONS}
In this study, we present a computational framework of using user-level multimedia posts from Instagram to investigate the effects of pets on happiness in terms of multiple aspects of user demographics. An Inception V3, which is retrained specifically on social media data, and timeline analysis together form the pet-owner classifier. The Face++ engine is employed for face detection, recognition, and user demographics inference. Among pet and none-pet owners, we also examine the happiness over different genders, different races, having partner or not, and having child or not. We believe our proposed framework is applicable to other related domains as a scalable, efficient, and effective methodology of social media user behavioral, and psychological well-being modeling and analysis.

\section{LIMITATION AND FUTURE WORK}
Our study infers user relationship status, and if having children by checking the posting frequency of faces, plus age differences. However, there are possible and perhaps rare exceptions. For instance, if a user frequently posts pictures with his/her close friends, this user maybe classified as having a partner (although arguably a close friend serves some of the social fucntions of a partner). It is also possible that a user frequently posts a child star, and this user could be identified as having a child. Nonetheless, we believe the trends and distributions in our large-scale study will remain consistent and valid, especially given that a significantly large number of users and images are analyzed.

Our work only includes data collected from Instagram, so this may not reflect the happiness level across different cultures outside US. Indeed, culture is one of the most important variables that impacts happiness \cite{Mathews2014}. In the future, we plan to collect multimedia posts through other countries' social media platforms, such as Chinese Wechat and Weibo, to analyze happiness in terms of culture - another important factor. Also in our current work, we consider the happiness scores derived from visual and textual contexts separately. It can be even more effective to infer individual's happiness if integrating those two types of scores organically. 

According to \cite{ura_2016}, many other factors, such as education, community and living standard, can impact individual happiness; those factors can be integrated into our happiness analysis if users education and employment records are accessible from social media such as LinkedIn. Furthermore, personality is also associated with the pursuit of happiness \cite{Tkach2006}, and may be used as a predictor of happiness \cite{Cheng2002}. According to \cite{Gottschalk1997}, words people use in their daily lives can provide rich information about their beliefs, fears, thinking patterns, social relationships, and personalities. By collecting user-level tweets plus leveraging Language and Inquiry and Word Count (LIWC2015) \cite{Pennebaker2015}, \cite{pan2017understanding} successfully connected the personality with career progression. In the same vein, we can also form an effective and automatic personality analysis tool for studying the effects of personality on happiness at a large scale. 

Furthermore, social multimedia posts offer the potential to provide a method for mass screening for individuals at risk for a range of poor health and psychological conditions. In particular, insufficient and low-quality sleep, unhealthy work-life balance are highly accountable for one's loss of happiness. \cite{peng2017sleep} and \cite{peng2017large} used user-level multimedia posts to study sleep deprivation and human fatigue in a massive scale. With some extension, the correlation between sleep, fatigue and happiness can also be quantified. In addition, Instagram has been used to pick up signals when individuals use language that could be related to risky behaviors \cite{Pang2015}. In the same way, our proposed framework can be extended to an automatic tool to monitor individual happiness to detect risk such as depressions, and allow for early intervention. 

\begin{acks}
We thank the support of New York State through the Goergen Institute for Data Science, our corporate research sponsors Xerox and VisualDX, and NSF Award \#1704309.  
\end{acks}

\bibliographystyle{ACM-Reference-Format}
\bibliography{bibliography}

\end{document}